\documentclass[aps, superscriptaddress, nofootinbib,  twocolumn]{revtex4}
\usepackage{amsmath,amssymb,bm,natbib}
\usepackage{epsfig}
\usepackage{graphicx}
\usepackage{amsmath}
\usepackage{color}

\newcommand{\be}{\begin{equation}}
\newcommand{\ee}{\end{equation}}
\newcommand{\beq}{\begin{equation}}
\newcommand{\eeq}{\end{equation}}
\newcommand{\bea}{\begin{eqnarray}}
\newcommand{\eea}{\end{eqnarray}}

\newcommand{\ie}{{\it i.e.}}

\newcommand{\ansatz}{{\it ansatz}}

\let\ifcomments\iftrue
\def\commentsoff{\global\let\ifcomments\iffalse}
\let\commentsize\small
\def\tinycomments{\global\let\commentsize\footnotesize}

\begin{document}

\vspace{3.0cm}

\title{Functional-analysis based tool for testing quark-hadron duality}
\author{Irinel Caprini}
\affiliation{Horia Hulubei National Institute for Physics and Nuclear Engineering, POB MG-6, 077125 Bucharest-Magurele, Romania}
\author{Maarten Golterman}
\affiliation{Department of Physics and Astronomy, San Francisco State
University, 1600 Holloway Ave, San Francisco, CA 94132, USA}
\author{Santiago Peris}
\affiliation{Department of Physics, Universitat
Aut{\`o}noma de Barcelona, 08193 Barcelona, Spain}
\vspace{3.0cm}

\begin{abstract} Quark-hadron duality is a key concept in QCD, allowing for the description of physical hadronic observables in terms of quark-gluon degrees of freedom. The modern theoretical framework for its implementation is Wilson's operator product expansion (OPE), supplemented by analytic extrapolation from large Euclidean momenta, where the OPE is defined, to the Minkowski  axis, where observable quantities are defined.  Recently, the importance of additional  terms in the expansion of QCD correlators
near the Minkowski axis,  responsible for quark-hadron duality violations (DVs), was emphasized. 
In this paper we introduce a mathematical tool that might be useful for the study of DVs in QCD. It is based on finding the minimal  distance, measured in the $L^\infty$ norm along a contour in the complex momentum plane, between a class of admissible functions containing the physical amplitude and the asymptotic expansion predicted by the OPE. This minimal distance is given by the norm of a Hankel matrix that can be calculated exactly, using as input the experimental spectral function on a finite interval of the timelike axis. We also comment on the relation
between the new functional tool and the more commonly used $\chi^2$-based analysis.
The approach is illustrated on a toy model for the QCD polarization function recently proposed in the literature.
\end{abstract}
\maketitle
\section{Introduction}

The application of theoretical predictions calculated in QCD in terms of  quarks and gluons  to measured hadronic observables is not straightforward and is usually based on assuming
the validity of ``quark-hadron duality." This concept, first used in a rather vague sense,  became more precise after its  modern  implementation in the framework of Wilson's operator product expansion (OPE) \cite{Shif, Blok, Shif1}.   Quark-hadron duality assumes that the description
in terms of the OPE, valid away from the Minkowski axis, can be analytically continued to
match with the description in terms of hadrons, which live on the Minkowski axis.\footnote {For 
an early discussion of quark-hadron duality, see Ref.~\cite{PQW}.}

In reality,  quark-hadron duality is violated.
This  was discussed first in detail in Refs.~\cite{Shif, Blok, Shif1, GoPe, Cata1, Cata2, Cata3}  and more recently in Refs.~\cite{Boito1, Boito2, Jamin1, Jamin2, Jamin4}.
The violation of duality implies the presence of new terms, in addition to the standard OPE of QCD Green functions, which are necessary in order to obtain agreement with experiment.  In principle, the necessity of such terms is not a surprise: since one tries to describe a function known to satisfy a  dispersion relation with hadronic unitarity thresholds within a framework  that uses quarks and gluons as degrees of freedom, the description cannot be exact, especially near the timelike axis in the complex momentum plane.  The question is how large these terms are, and whether they can be calculated theoretically.  Of course, one expects the duality violating terms to be 
 small in the Euclidean region, far  from the unitarity cut, and larger near the timelike
axis. Unfortunately, no theoretical calculations of duality violations (DVs) are available.
Therefore, in order to obtain insight into these terms one must resort to phenomenological models. However, the phenomenological extraction of DVs is made difficult by the fact that both 
the perturbative expansion in QCD as well as the power-suppressed terms in the OPE are 
expected to be divergent series.
So, as long as we deal with truncated series, it is difficult
to disentangle genuinely new terms predicted by specific dynamical mechanisms \cite{Shif, Shif1, Cata1} from contributions of the higher-order terms in the standard QCD expansions, which can be quite large since the expansions are divergent. 
 
For our discussion it is crucial to observe that the analyticity properties in the complex 
momentum plane of the
QCD correlation functions calculated with the OPE are different from those of the exact functions.
As for example discussed in Ref.~\cite {Oehme}, in the confined phase of QCD 
the poles and the branch-points of the true Green functions are generated by
 physical hadron states, and no singularities 
 related to the underlying quark and gluon degrees of freedom
should appear. On the other hand, in perturbation theory the branch-points 
of the Green's functions are produced by the quarks and gluons appearing in 
Feynman  diagrams. Other, more complicated singularities
make their appearance if one goes beyond simple perturbation theory.
For example, renormalization group improved expansions have unphysical
spacelike
singularities (Landau poles), and power corrections   involving nonperturbative vacuum condensates  exhibit poles at the origin \cite{SVZ}.
 Moreover, partial resummations of particular terms in the perturbative expansion,
like for instance renormalon chains, might introduce other
singularities  \cite{CaNe}. For a general review of renormalon singularities, see
Ref.~\cite{MB}.

The method we develop in this paper distinguishes between the exact 
 Green's function (which  has
 the correct analyticity properties), for which we have
incomplete information available from experiment, and the approximant provided by theory,  which has different analytical properties.
As we shall show below, the difference between these two functions, 
 measured in a suitable norm
along a contour in the complex momentum plane,
must be larger than (or equal to) a certain nonnegative number $\delta_0$, which can be computed by techniques of functional analysis.   The value of $\delta_0$ 
can then be used to distinguish between different theoretical approximants, such as
for instance approximants involving only the OPE, and approximants involving the OPE and
a model for DVs. A smaller value of $\delta_0$ indicates a better approximant.

The paper is organized as follows: in the next section, starting from the information available about a generic polarization function in QCD, we formulate a suitable functional optimization problem. It leads us to consider the minimum distance, $\delta_0$, measured in the $L^\infty$ norm along a contour in the complex momentum plane, between the QCD approximant and the class of admissible functions containing the physical amplitude. The solution of the problem is presented in Sec.~\ref{sec:solution}. 
In  Sec.~\ref{sec:model}  we discuss the properties of $\delta_0$ in the context of a model for the vector polarization function, proposed in Ref.~\cite{Cata2}. In Sec.~\ref{sec:dv} we consider in particular   the dependence of  $\delta_0$ on the strength of the DVs in the same model. In Sec.~\ref{sec:disc}, starting from the solution of the functional problem derived in this paper,  we discuss a test based on the $L^\infty$ norm, as a possibly interesting alternative to 
 the usual $\chi^2$ (which is based on the $L^2$ norm) for discriminating between different DV models. In Sec.~\ref{sec:out}  we present our conclusions.  
 
 \section{Formulation of the problem}\label{sec:formulation}
We consider a  light-quark vector vacuum polarization function $\Pi(s)$ in QCD. 
From causality and unitarity it is known that 
the  function $\Pi(s)$ is  analytic of real type (\ie, $\Pi (s^*)=[\Pi (s)]^*$) in the
complex  $s$-plane cut along the positive
 real axis above the lowest threshold ($4m_\pi^2$)
for hadron production.
These properties are implemented in the once subtracted
dispersion relation of the form
\begin{equation}\label{disprel}
\Pi(s)= \Pi(0)+ \frac{s}{ \pi}\int_{4m_\pi^2}^\infty \frac{\sigma(s')
}{ s'(s'-s-i\epsilon)}\, d s',
\end{equation}
where the limit $\epsilon\to 0^+$ is always understood.
We will assume that the spectral function 
\begin{equation}\label{sigma}
\sigma(s)\equiv \mbox{Im}\,\Pi(s+i\epsilon)
\end{equation}
 is known  experimentally on a limited energy range,  $4m_\pi^2\leq s\leq s_0$.  
For instance, the spectral functions of both vector and axial currents of light quarks are measured in the hadronic $\tau$ decays up to $s_0= m_\tau^2$ \cite{Aleph, Aleph1, Opal}.   For the use of 
hadronic $\tau$ decays for the determination of the strong coupling $\alpha_s$, see Refs.~\cite{Aleph, Aleph1, Opal,BrNaPi, DiPi, Davier2008, MaYa, BeJa,  CaFi2009, Pich2010,Boito1, CaFi2011, Boito2, AACF,BBJ, AACF1,BCK08}.

\begin{figure}[t]
\vspace{0.3cm}
\includegraphics[width=2.8in]{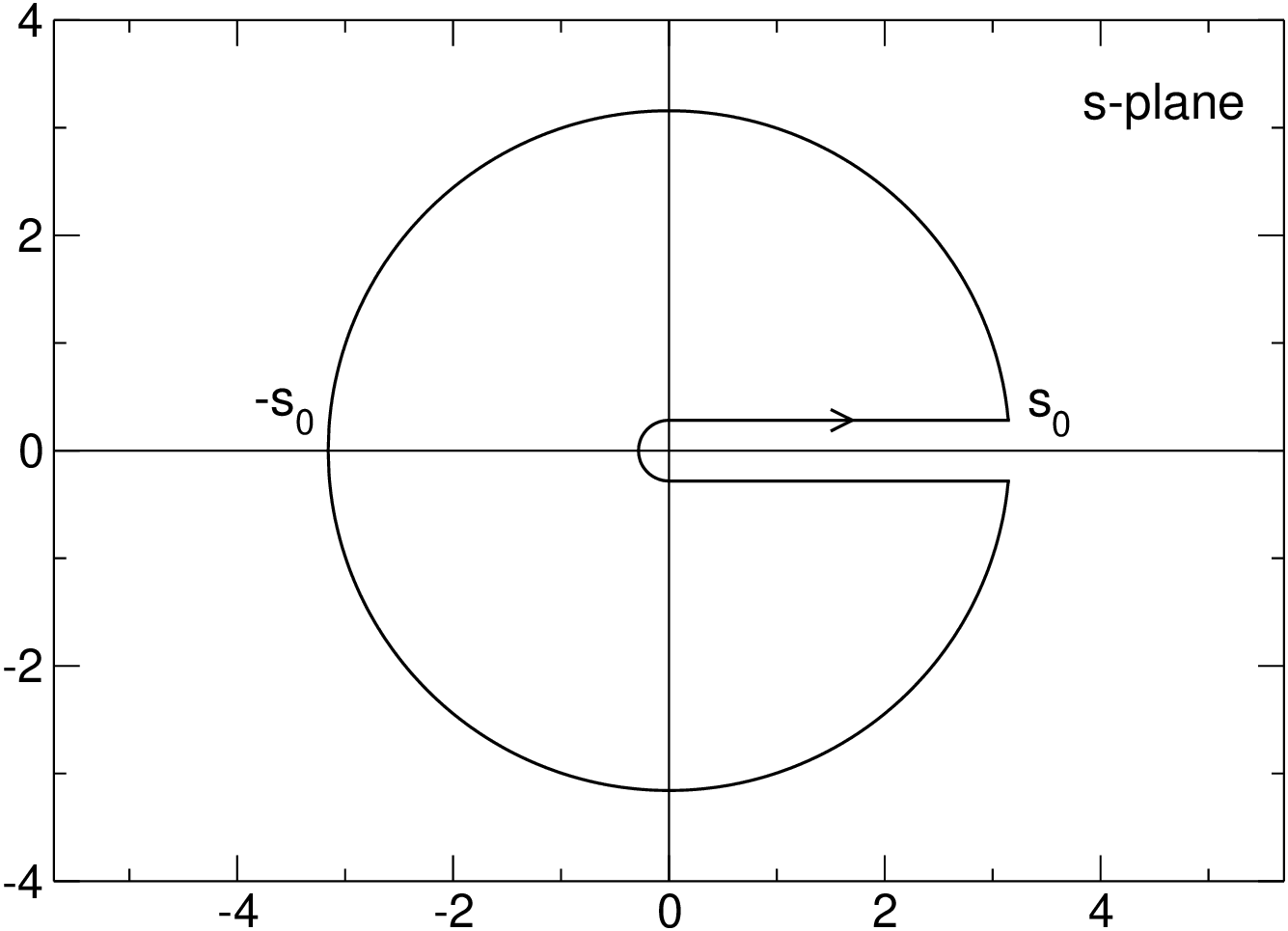}
\caption{Analytic properties of $\Pi(s)$ in the disk $|s|<s_0$ of the complex $s=q^2$ plane.  $\Pi(s)$ is analytic for all $s$ except $4m_\pi^2\le s<\infty$ on the positive real axis.
\label{fig:0}}
\end{figure}

At large momenta sufficiently away from the timelike cut, $\Pi(s)$ can be calculated in the framework of the OPE, in terms of nonperturbative condensates and perturbative
coefficients \cite{SVZ}. As discussed recently \cite{Shif, Blok, Shif1, Cata1, Cata2, Cata3}, additional terms are required beyond the OPE in order to improve the description near the timelike axis. Including these  duality violating terms, the QCD prediction is written as
 \beq\label{eq:OPE}
\Pi_{\rm QCD}(s)=\Pi_{\rm OPE}(s)+\Pi_{\rm DV}(s),
\eeq
where the OPE contribution can be separated into a pure perturbative (dimension $D=0$) part and the power-corrections (PC):
 \beq\label{eq:pertPC}
\Pi_{\rm OPE}(s)=\Pi_{\rm pert}(s)+ \Pi_{\rm PC}(s).
\eeq
 As discussed above, the exact 
function $\Pi(s)$ cannot coincide with  $\Pi_{\rm QCD}(s)$ even after suitable resummations of the perturbative part.  In order to quantify this deviation,  we consider the maximum value of the difference between the two functions along a contour $\Gamma$  in the
complex $s$ plane, which
intersects the real positive axis at $s=s_0$.  In particular,  if  $\Gamma$ is the circle $s=
s_0 e^{i\theta}$, which is part of the contour shown in Fig.~\ref{fig:0}, we consider the difference
\begin{equation}\label{diff}
\delta = \sup_{\theta\in (0,2\pi)}\vert \Pi(s_0 e^{i\theta})-\Pi_{\rm QCD}(s_0 e^{i\theta})\vert,
\end{equation}
where $0\le \theta\le 2 \pi$ denotes the angle of the ray in the complex plane (the upper and lower edges of the unitarity cut corresponding to $\theta = 0$ and $\theta = 2 \pi$, respectively). 
Even though the exact function $\Pi(s_0 e^{i\theta})$ is not known,
 we shall prove in the next section
that the quantity $ \delta$ defined above  must be greater than or equal to a certain
 nonnegative quantity which can be explicitly calculated.\footnote{Similar functional-analysis methods were applied for scattering amplitudes in particle physics in Refs.~\cite{CaCi}, while in the context of QCD such techniques have been applied in Refs.~\cite{CaVe}.}

In order to write the problem in a canonical form, we first make a conformal mapping of the contour $\Gamma$  onto the  contour $\vert z\vert=1$ of 
the plane $z\equiv \tilde z(s)$, where  $\tilde z(0)=0$ and  $\tilde z(s_0)=1$. For the circle shown in Fig.~\ref{fig:0},  the  transformation is trivial,
\begin{equation}\label{z}
\tilde z(s)=\frac{s}{s_0}\,.
\end{equation}
Using this change of variable in Eq.~(\ref{disprel}),
 the  function $\Pi(s)$ can be written as a function of $z$ in the form
\begin{equation}\label{eq:Piz}
\Pi(s_0 z)=\frac{z}{\pi}\int_{x_0}^{1+\eta}dx\,\frac{\sigma(s_0 x)}{x(x-z-i\epsilon)} + g(z)\,,
\end{equation}
with $x_0=4m_\pi^2 /s_0$, and where $g(z)$ is an unknown function, real analytic  in the unit disk  $|z|\le 1$.
Then, if we define the function $h(\zeta)$  for $\zeta=\exp(i\theta)$ as
\beq\label{h}
h(\zeta)= -\frac{\zeta}{\pi}\int_{x_0}^{1+\eta}dx\,\frac{\sigma(s_0 x)}{x(x-\zeta-i\epsilon)} + \Pi_{\rm QCD}(s_0 \zeta)\,,
\eeq 
$h$ is bounded on the circle $|\zeta|=1$,\footnote{This is the reason we introduced the
positive number $\eta$ in Eq.~(\ref{eq:Piz}).} and
Eq.~(\ref{diff}) can be written as
\beq\label{eq:delta}
\delta = \Vert g-h\Vert_{L^\infty}\,,
\eeq
where we introduced the $L^\infty$ norm of a function, defined as the supremum of the modulus
along the boundary $\vert z\vert=1$: 
\begin{equation}\label{Hinf}
\Vert F\Vert_{L^\infty}\equiv \sup_{\theta\in (0,2\pi)}\vert F(e^{i\theta}) \vert.
\end{equation} 
The norm $\delta$ does not depend on the value of the positive number $\eta$,
because a variation in its value gets absorbed into a change of the function $g$.
 
 Starting from Eq.~(\ref{eq:delta}), we now consider the following functional extremal problem: find
\begin{equation}\label{inf1}
\delta_0=\min_{g\in H^\infty} \Vert g-h\Vert_{L^\infty}\le\delta\,,
\end{equation}
where  the minimization is with respect to all the functions $g(z)$
analytic  in the disk $\vert z \vert<1$ and bounded on its boundary (this class of functions is  denoted as $H^\infty$ \cite{Duren}).

The quantity $\delta_0$ provides the lower bound announced below Eq. (\ref{diff}). In the next section we present the solution of the problem~(\ref{inf1}) and the algorithm for calculating $\delta_0$. 

\section{Solution of the extremal problem}\label{sec:solution}
We 
shall apply a ``duality theorem" in functional  optimization theory
\cite{Duren}, which  replaces the original minimization problem~(\ref{inf1})
by a maximization problem in a different functional space, denoted as the
``dual" space. The new problem will turn out to be easier than the original one, 
allowing us to obtain the solution by means of a numerical algorithm. 

Let us denote by $H^p$, $p<\infty$, the class of functions $F(z)$ which are
analytic inside the unit disk $|z|<1$ and satisfy the boundary condition
\begin{equation}\label{Hp}
\|F\|_{L^p}\equiv \left[\frac{1}{ 2\pi}\int_0^{2\pi}\vert F(e^{i\theta})\vert^p d\theta\right]^{1/p} <\infty.
\end{equation}
In particular, $H^1$ is the Banach space of analytic functions
with integrable modulus on the boundary, and $H^2$ is the Hilbert space
of analytic functions with integrable modulus squared. We introduced already
  the class  $H^\infty$ of functions analytic and bounded in
 $\vert z\vert <1$, for which the $L^\infty$ norm was defined
in Eq.~(\ref{Hinf}). The classes $H^p$ and $H^q$ are said to be dual if the relation $1/p+1/q=1$ holds \cite{Duren}. It follows that $H^1$ and $H^\infty$ are dual to each other, while $H^2$ is dual to itself.

We now state the duality theorem of interest to us. 
Let $h(\zeta=\mbox{exp}(i\theta))$ be an element of $L^\infty$.
Then the following equality holds (see Sec.~8.1 of Ref.~\cite{Duren}):
\begin{equation}\label{dual}
\min_{g\in H^\infty}\|g-h\|_{L^\infty}=\sup_{G\in S^1}
\frac{1}{2\pi}\left\vert\oint_{\vert \zeta\vert =1} G(\zeta)h(\zeta) d \zeta
\right\vert,
\end{equation}
where $S^1$ denotes the unit sphere of the Banach space $H^1$, \ie, the set of functions 
$G\in H^1$ which satisfy the condition $ \|G\|_{L^1}\leq 1$.

We note first that the equality~(\ref{dual}) is automatically satisfied  if $h(\zeta)$
 is the boundary value of an analytic 
function in the unit disk, since in this case
the minimal norm on the left-hand side is zero, and the right-hand side of Eq.~(\ref{dual}) vanishes too, by Cauchy's theorem.
 The nontrivial case corresponds  to a function $h(\zeta)$ 
 which  is not the
boundary value of a function analytic in
 $|z|< 1$. In that case, $h(\zeta)$ will admit the general Fourier expansion
\beq\label{hpm}
h(\zeta)=\sum\limits_{n=0}^\infty h_n \zeta^n + \sum\limits_{n=1}^\infty c_n \zeta^{-n}\,, 
\eeq
where, due to reality property of $\Pi(s)$ mentioned above, the coefficients $h_n$ and $c_n$ are real.{\footnote{This property will be valid for the coefficients of all expansions below.} The analytic continuation of the expansion (\ref{hpm})  inside $|z|<1$ will contain both an analytic part (the first sum) and a nonanalytic part (the second sum). Intuitively, we expect the minimum norm 
in Eq.~(\ref{dual}) to depend explicitly only on the nonanalytic part, \ie, on the coefficients $c_n$. The proof given below will confirm this expectation.

In order to evaluate the supremum on the right-hand side of Eq.~(\ref{dual}) we
 use  a factorization theorem (see the proof of Theorem~3.15 in Ref.~\cite{Duren}) according to which
every function $G(z)$ belonging to the unit sphere $S^1$ of $H^1$ 
 can be written as
\begin{equation}\label{factor}
G(z)=w(z)f(z)\,,
\end{equation}
where the functions $w(z)$ and $f(z)$ belong to the unit sphere $S^2$ of $H^2$,
\ie, are analytic and satisfy the conditions
\begin{equation}\label{wGnorm}
\Vert w\Vert_{L^2} \leq 1\,,\quad \quad\quad\Vert f\Vert_{L^2}\leq 1\,.
\end{equation}
 Therefore, if one writes the Taylor expansions
 \begin{equation}\label{wf}
w(z)=\sum_{n=0}^\infty w_nz^n\,,\quad \quad f(z)= \sum_{m=0}^\infty f_m z^m\,,
\end{equation}
the coefficients satisfy  the conditions
\begin{equation}\label{wfl2}
\sum_{n=0}^\infty w_n^2\leq 1\,,\quad\quad \quad\sum_{m=0}^\infty f_m^2\leq 1\,.
\end{equation} 
After introducing the representation~(\ref{factor}) into Eq.~(\ref{dual}), we obtain the equivalent  relation
\begin{equation}\label{dual1}
\min_{g\in H^\infty}\|g-h\|_{L^\infty}=\sup_{w,f \in S^2}\left\vert \frac{1}{ 2\pi}
\oint\limits_{\vert \zeta\vert =1} w(\zeta)f(\zeta)h(\zeta) d\zeta\right\vert\,,
\end{equation}
where the supremum on the right-hand side is taken with respect to the functions 
$w$ and $f$ with
the properties mentioned in Eqs.~(\ref{wf}) and (\ref{wfl2}). By inserting
 into  Eq.~(\ref{dual1}) 
the expansions (\ref{wf})
we obtain, after a straightforward calculation
\begin{equation}\label{dual2}
\min_{g\in H^\infty}\|g-h\|_{L^\infty}=\sup_{\{w_n,f_m\}}
\left\vert\sum_{m,n=1}^\infty {\cal H}_{nm}w_{n-1}f_{m-1}\right\vert\,.
\end{equation}
Here the supremum is taken with respect to the sequences 
$w_n$ and $f_m$ subject to the condition (\ref{wfl2}), and the numbers
\begin{equation}\label{hank}
{\cal H}_{nm}=c_{n+m-1}\,, \quad n,m\geq 1\,,
\end{equation}
define a matrix\footnote{Matrices with elements defined in this way are called Hankel matrices \cite{Duren}.} ${\cal H}$ in terms of the  negative frequency Fourier coefficients $c_n$ of the function $h$ expanded in Eq.~(\ref{hpm}), which are calculated as
\begin{equation}\label{four}
c_{n}=\frac{1}{2\pi }\int\limits_0^{2 \pi} e^{i n\theta}h(e^{i\theta})  d\theta =\frac{1}{2\pi i}\oint\limits_{\vert \zeta\vert=1}\zeta^{n-1}h(\zeta) d\zeta\,.
\end{equation} 
Thinking of $w_{n-1}$ and $\sum_m{\cal H}_{nm}f_{m-1}$ as the components of vectors
${\bf w}$ and ${\bf {\cal H}f}$,  the
absolute value of the sum in Eq.~(\ref{dual2}) can be written as
$\left\vert{\bf w}\cdot{\bf {\cal H}f}\right\vert$, and the Cauchy--Schwarz inequality implies that it satisfies
\beq
\label{CS}
\left\vert{\bf w}\cdot{\bf {\cal H}f}\right\vert\le\Vert{\bf w}\Vert_{L^2}
\Vert{\bf {\cal H}f}\Vert_{L^2}\le\Vert{\bf {\cal H}f}\Vert_{L^2}\,.
\eeq
Since Eq.~(\ref{CS}) is saturated for ${\bf w}\propto{\bf {\cal H}f}$, it follows that the
supremum in Eq.~(\ref{dual2}) is given by the $L^2$ norm of ${\bf {\cal H}}$.
The solution of the minimization problem~(\ref{inf1}) can then be written as 
\begin{equation}\label{delta0}
\delta_0=\Vert{\cal H}\Vert_{L^2}=\Vert{\cal H}\Vert\,,
\end{equation}
where $\Vert{\cal H}\Vert$ is the spectral norm, given by 
 the square root of the greatest
 eigenvalue of the positive-semidefinite matrix ${\cal H}^\dagger {\cal H}$.
 In numerical  calculations \cite{CaSa},  the matrix is truncated  at a finite order 
$m=n=N$,  and   the convergence of the successive approximants with increasing $N$ is checked.
 By the duality theorem, the initial
functional minimization problem~(\ref{inf1})
 was therefore reduced to a rather simple numerical computation. We finally note that, using the expression (\ref{h}) for the function $h$, the coefficients (\ref{four}) can be written as
\be\label{four1}
c_n=\frac{1}{\pi}\int\limits_0^{1} x^{n-1} \sigma(s_0 x) dx +\frac{1}{2\pi}\int\limits_0^{2 \pi} e^{i n\theta} \Pi_{\rm QCD}(s_0 e^{i \theta} )d\theta\,,
\ee
where in the first term we recognize the  moments of the spectral function, which in physical applications are known from experimental measurements.\footnote{Note that 
the first term in Eq.~(\ref{four1}) does not depend on $\eta$, because the contour integral in Eq.~(\ref{four}) yields
zero for $1<x<1+\eta$.}

For our further discussion it is of interest to consider also the minimization problem similar to
Eq.~(\ref{inf1}), where the $L^\infty$ norm is replaced by the $L^2$ norm
\begin{equation}\label{infL2}
\delta_2=\min_{g\in H^2} \Vert g-h\Vert_{L^2}\,.
\end{equation}
The solution of this problem is obtained easily by noting that the $L^2$ norm squared of the difference $g-h$ is 
\beq\label{infL21}
\Vert g-h\Vert_{L^2}^2=\sum\limits_{n= 0}^\infty(g_n-h_n)^2+\sum\limits_{n=1}^\infty c_n^2\,,
\eeq
where $g_n$ are the coefficients of the Taylor expansion at $z=0$ of the analytic function $g(z)$. The minimum of the right-hand side is reached for $g_n=h_n$, $n\ge 0$, and reduces to the second sum. Therefore,  the solution of the problem~(\ref{infL2}) is
\beq\label{delta2}
\delta_2 =[\sum_{n= 1}^\infty\ c_n^2\ ]^{1/2}\,.
\eeq
Using the fact that the $L^\infty$ norm of every function is larger than its  $L^2$ norm, one finds that
\beq\label{l2}
 \delta_0\ \ge \ \delta_2\,,
\eeq
where  the inequality is strict, except for the trivial case of a function $h(\zeta)$ whose expansion (\ref{hpm}) contains a single term in the nonanalytic part. We shall return to the relation between the quantities $ \delta_0$ and  $ \delta_2$ in Sec. \ref{sec:disc}.
\section{Application to a specific model}\label{sec:model}
We shall illustrate the properties of the quantity $\delta_0$ on a  model for the vector polarization function, proposed in Ref.~\cite{Cata2} (based on an idea in Refs.~\cite{Blok,Shif1}), where we can calculate everything exactly. We adopted in fact a simplified version of the model, obtained by including the $\rho$ pole 
into the ``Regge tower'' of resonances by adjusting the value of $m_0$, from Eq.~(3.7) of Ref.~\cite{Cata2}. Then $\Pi(s)$ is defined by the model function 
\begin{equation}\label{model0}
\Pi_{\rm model}(s)=-\frac{1}{\zeta}  \frac{2 F^2}{\Lambda^2}\
\psi\left(\frac{v+m_0^2}{\Lambda^2}\right)\,,
\end{equation}
in terms of the Euler digamma function $\psi(v)=\Gamma\,'(v)/\Gamma(v)$, where the variable $v$ is defined as 
\begin{equation}\label{v}
v=\Lambda^2\left(\frac{-s-i\epsilon}{\Lambda^2}\right)^{\zeta}\,,
\end{equation}
and the parameters have the numerical values
\begin{eqnarray}\label{param}
 \zeta &=& 0.95\,,\  F=133.8\ \mathrm{MeV}\,,\\
 \  \Lambda&=&1.189\ \mathrm{GeV}\,, \ m_0=0.75\ \mathrm{GeV}\,.\nonumber
\end{eqnarray}
The ``OPE expansion" of $\Pi_{\rm model}$ was obtained in Ref.~\cite{Cata2} using the asymptotic expansion of the digamma function, valid for $ |v|\gg 1,\ -\pi< \arg(v) < \pi$,
\beq\label{psias}
\psi(v)=\log{v}-\frac{1}{2v}-\sum_{n=1}^{\infty}\frac{B_{2n}}{2n\, v^{2n}}\,,
\eeq
 where $B_{2n}$ are the Bernoulli numbers.  Although the  expansion~(\ref{psias}) converges for no $v$, truncated sums provide a good approximation for large $|v|$. From Eq.~(\ref{psias}) one can derive the expansions of $\Pi_{\rm model}$ truncated  at a finite order $N_{\rm OPE}$ as \cite{Cata2}
\beq\label{opemodel}
\Pi_{\rm OPE}(s) =-\ \frac{2 F^2}{\Lambda^2}\, C_0 
\log\left(\frac{-s}{\Lambda^2}\right)\ +\sum_{k=1}^{N_{\rm OPE}}\frac{C_{2k}}{v^k}\ ,
\eeq
where the first term represents the ``purely perturbative'' part $\Pi_{\rm pert}$ and the other ones are the higher-dimensional (PC) terms. From Ref.~\cite{Cata2}
\begin{equation}\label{condensate}
C_0=1\,, \quad  C_{2k}=\frac{2}{\zeta} (-1)^k \frac{1}{k}
\Lambda^{2k-2} F^2 B_k\left(\frac{m_0^2}{\Lambda^2}\right),
\end{equation}
where $B_k\left(x\right)$ stand for the Bernoulli polynomials. 

As in QCD, the  OPE expansion~(\ref{opemodel}) is not accurate near the timelike axis, where DVs are expected to have a large effect. The description was improved by taking into account the reflection property
\begin{equation}\label{reflection}
    \psi(v)=\psi(-v)-\pi\cot{(\pi v)}-\frac{1}{v}\,,
\end{equation}
which suggests a modified approximant for $\Pi_{\rm model}(s)$, valid for large values of $|s|$ and ${\rm Re}\ s>0$
\begin{equation}\label{OPEDV} \Pi_{\rm model}(s)\approx \Pi_{\rm OPE+DV}(s) \equiv\, 
\Pi_{\rm OPE}(s)+\Pi_{\rm DV}(s)\,.
\end{equation}
The correction $\Pi_{\rm DV}(s)$  is given in the upper half-plane ${\rm Im}\ s>0$ of the right half-plane $\mathrm{Re}\ s > 0$ by the expression
\begin{equation}\label{PiDV}
\Pi_{\rm DV}(s)=\frac{2 \pi F^2}{\Lambda^2\zeta}\left[-i+
\cot\left[\pi\left(\frac{-s}{\Lambda^2}\right)^{\zeta}+\pi\,\frac{m_0^2}{\Lambda^2}\right]\right]\,, 
\end{equation}
and can be defined by means of the reality property $\Pi_{\rm DV}(s^*)=\Pi^*_{\rm DV}(s)$ in the lower half-plane. In the left half $s$-plane the correction is assumed to vanish, $\Pi_{\rm DV}(s)=0$ for $\mathrm{Re}\,s\leq 0$.

\begin{figure}[t]
\centering
\vspace{0.3cm}
\includegraphics[width=2.7in]{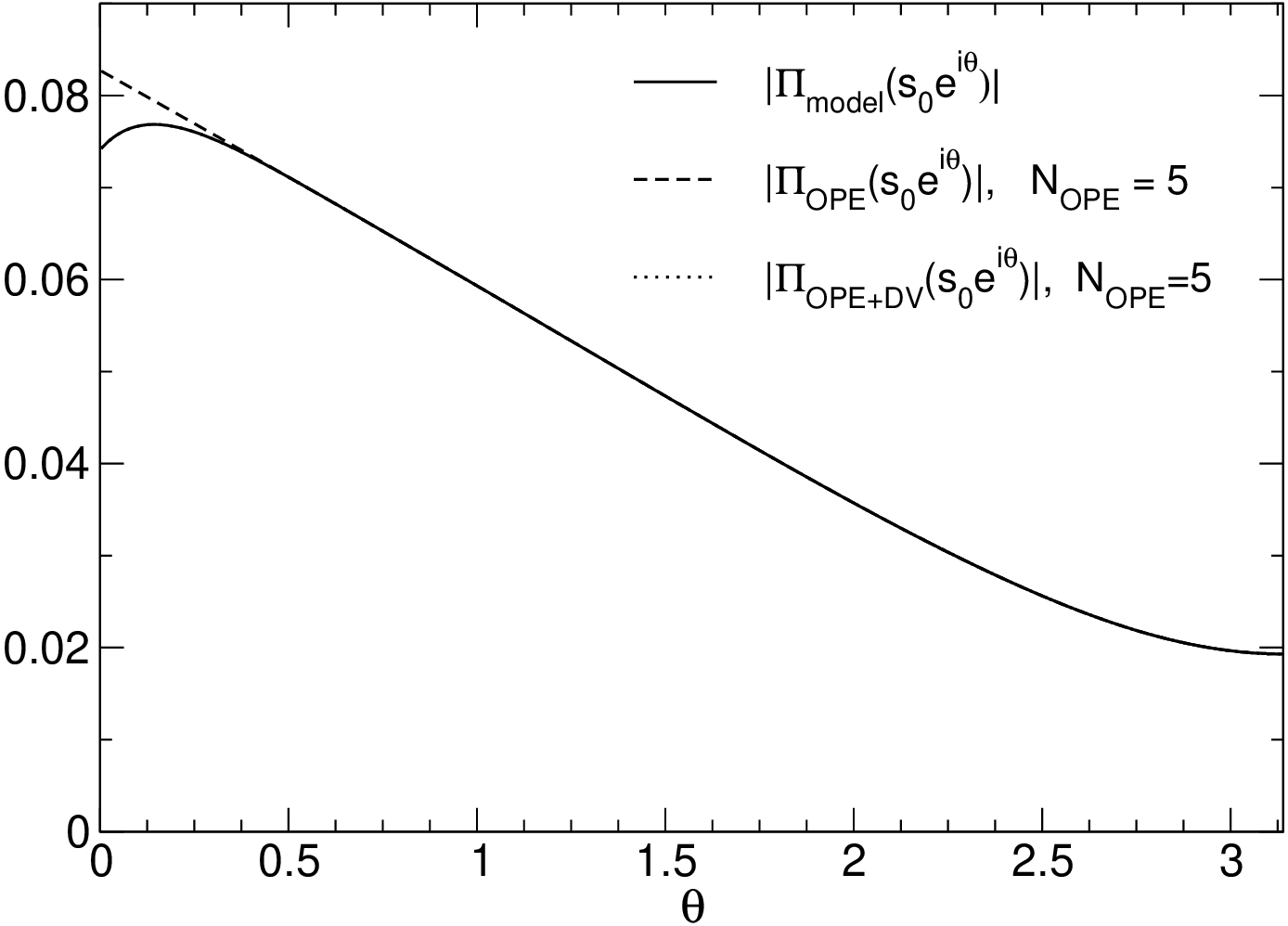}
\caption{Modulus of $\Pi_{\rm model}(s)$ defined in Eq.~(\ref{model0}) and of the OPE approximant~(\ref{opemodel}) truncated at $N_{\rm OPE}=5$, along the upper semicircle $s=s_0 e^{i\theta}$ with $s_0=m_\tau^2$. The OPE+DV approximant~(\ref{OPEDV})  is not visibly different from the exact function. \label{fig:1}}\vspace{0.3cm}
\end{figure}

\begin{figure}[b]
\centering
\vspace{0.3cm}
\includegraphics[width=2.7in]{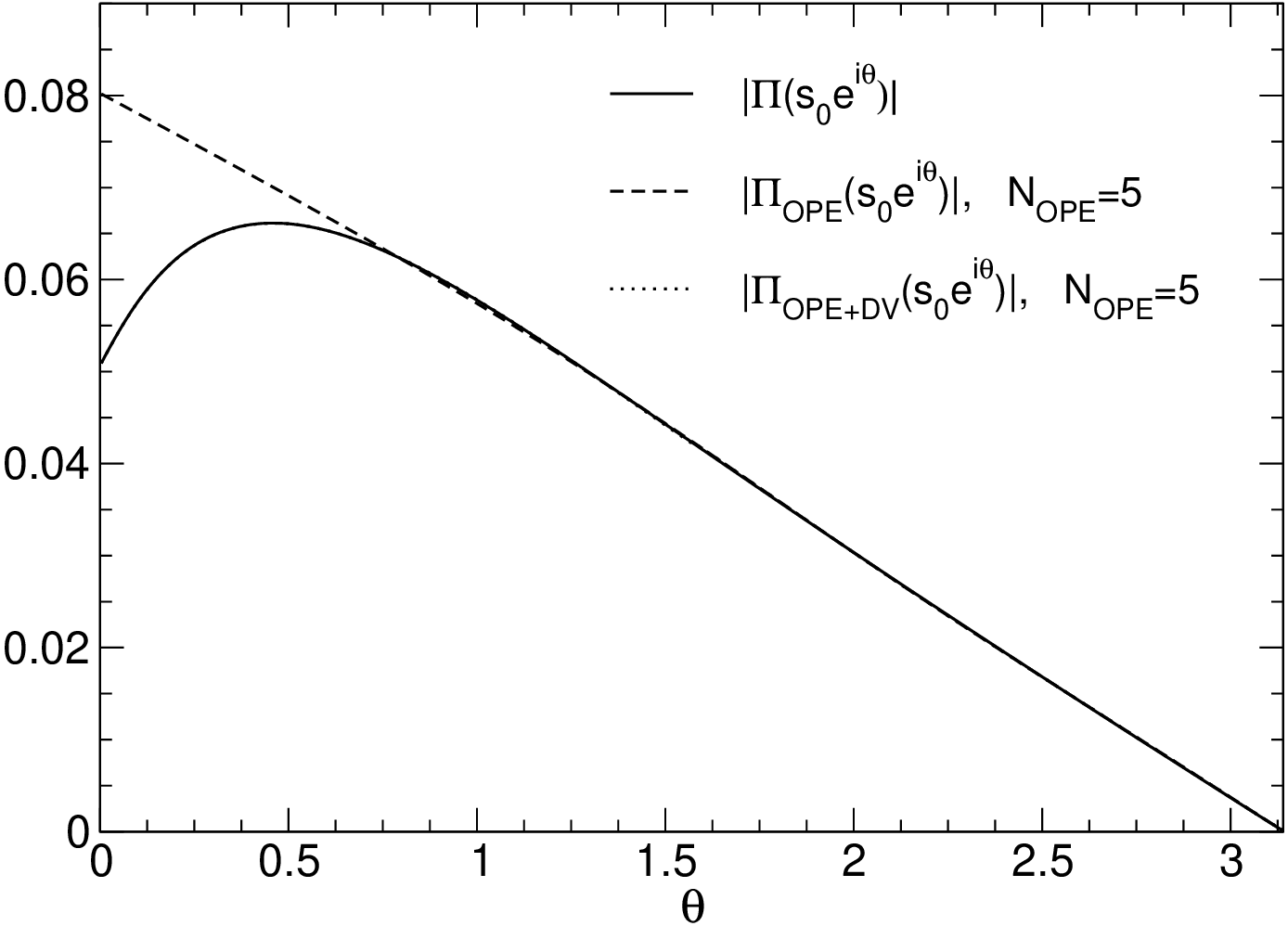}
\caption{As in Fig.~\ref{fig:1} for  $s_0=1.5\ {\rm GeV}^2$.  \label{fig:2}}\vspace{0.3cm}
\end{figure}

In order to illustrate the quality of the approximation, we show in Fig.~\ref{fig:1} the modulus of the exact function and of its approximants along the upper semicircle $s=s_0 e^{i\theta}$, $\theta\in (0, \pi)$, for $s_0=m_\tau^2$, for the truncation order $N_{\rm OPE}=5$. One can see that the OPE expansion provides a good approximation along the circle except close to the timelike axis, which corresponds to $\theta=0$. By adding the DV term, the approximant gets very close to the exact function $\Pi_{\rm model}(s)$,  and cannot be distinguished in the figure. 

For lower values of $s_0$ we expect the approximation to deteriorate progressively. For illustration, we show  in Fig.~\ref{fig:2} the modulus of the exact function and of its approximants along the circle of radius $s_0$ equal to $1.5\ {\rm GeV}^2$. The deviation between the exact function and the OPE near the timelike axis is larger, but the DV term restores the agreement also in this case.  

\begin{figure}[t]
\centering
\vspace{0.3cm}
\includegraphics[width=2.5in]{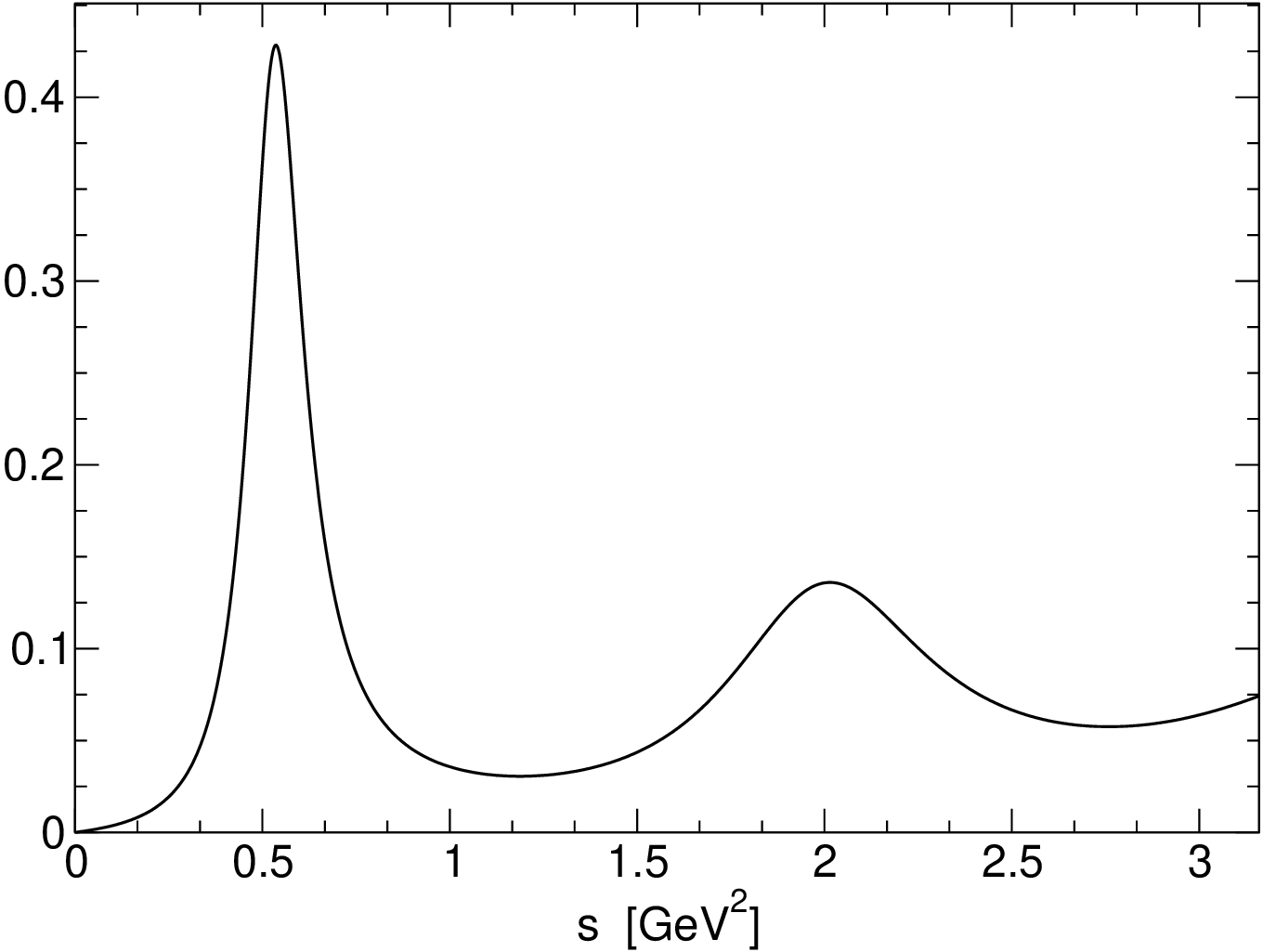}
\caption{Spectral function $\sigma_{\rm model}(s)$ for $s\leq m_\tau^2$. \label{fig:sigma}}\vspace{0.3cm}
\end{figure}

We will now use the model described above to mimic a typical situation in practical applications to QCD. In the region $s\leq s_0$ we will use as input the spectral function 
\beq\label{sigmamodel}
\sigma_{\rm model}(s) = \mbox{Im}\,\Pi_{\rm model}(s+i\epsilon)\,, \quad\quad s\leq s_0\,,
\eeq
calculated from the exact model (we show for illustration in Fig.~\ref{fig:sigma} this spectral function in the range from 0 to $m_\tau^2$).
Along the circle $|s|=s_0$ we use as input  the asymptotic expressions $\Pi_{\rm OPE}(s)$ or  $\Pi_{\rm OPE+DV}(s)$ defined in Eqs.~(\ref{opemodel}) and (\ref{OPEDV}).  Then the Fourier coefficients $c_n$ from Eq.~(\ref{four1}) have the specific form
\beq\label{cntot}
c_n=c_{n}^{\rm \sigma}+ c_{n}^{\rm OPE}+ c_{n}^{\rm DV}\,,\quad\quad n\ge 1\,,
\eeq
where
\bea\label{cnsigma}
c_{n}^{\sigma}&=&\frac{1}{\pi}\int\limits_0^{1} x^{n-1} \sigma_{\rm model}(s_0 x) dx\,,\\
c_n^{\rm OPE}&=&\frac{1}{2\pi }\int\limits_0^{2 \pi} e^{i n\theta}\Pi_{\rm OPE}(s_0 e^{i\theta})  d\theta \,,\\
c_n^{\rm DV}&=&\frac{1}{2\pi }\int_{\cal R} e^{i n\theta}\Pi_{\rm DV}(s_0 e^{i\theta})  d\theta\,.
\eea
Here  ${\cal R}$  is  the right semicircle, $\theta\in (0,\pi/2]\cup[3\pi/2,2\pi)$. 
In practice, due to the reality property of the integrands, to obtain  $c_n^{\rm OPE}$  and $c_n^{\rm DV}$ it is enough to integrate  only along the upper semicircle and take twice the real part. 
 Having the coefficients $c_n$, we computed with Mathematica the norm of the Hankel matrix~(\ref{hank}), truncated at a finite order $N$. The convergence was tested by increasing $N$ from 20 to 700. The results presented below are obtained with $N=500$.

In Table~\ref{tab:d0} we give the values of $\delta_0$ for the OPE expansion and the full OPE+DV approximant as a function of the truncation order $N_{\rm OPE}$ appearing in Eq.~(\ref{opemodel}).  We took $s_0=m_\tau^2$.  For comparison we also give the  difference
\beq\label{eq:dexactOPE}
\delta^{\rm OPE}_{\rm exact}= \sup_{\theta\in (0,2\pi)}|\Pi_{\rm model}(s_0 e^{i\theta})- \Pi_{\rm OPE}(s_0 e^{i\theta})|\eeq
between the exact function~(\ref{model0}) and the OPE expansion~(\ref{opemodel}), and the difference
\beq\label{eq:dexactOPEDV}
\delta^{\rm OPE+DV}_{\rm exact}= \sup_{\theta\in (0,2\pi)}|\Pi_{\rm model}(s_0 e^{i\theta})- \Pi_{\rm OPE+DV}(s_0 e^{i\theta})|
\eeq
between the exact function and the OPE+DV expression~(\ref{OPEDV}).
 For all cases shown in Table~\ref{tab:d0}, the inequality  $\delta_0 < \delta_{\rm exact}$ holds. This is what we expect, in fact: indeed, $\delta_0$ is the minimal value calculated over a class of functions to which the exact function $\Pi_{\rm model}(s)$ is supposed to belong. One may note that the values of 
 $\delta_{\rm exact}$ are not much larger that the lower bound $\delta_0$ in this case.  
 
In the table, we included the value of $N_{\rm OPE}=13$ for which 
$\delta^{\rm OPE+DV}_0$ and, simultaneously, $\delta^{\rm OPE+DV}_{\rm exact}$,
take on their minimal values. Actually, it is around
this order in the expansion of our model that the OPE starts diverging, for $s_0=m_\tau^2$.
Of course, the fact that the OPE is an asymptotic series  has been built into the model.
Note that the divergence of the OPE is not visible in $\delta^{\rm OPE}_0$ or
$\delta^{\rm OPE}_{\rm exact}$ (left half of the table) until much larger values of $N_{\rm OPE}$.
These quantities are less sensitive to the OPE because of the missing DV contribution.
 
\begin{table}[hbtp]
\centering
\caption{Values of $\delta_{\rm exact}$ and the  lower bound $\delta_0$ computed for the approximants $\Pi_{\rm OPE}(s)$ and $\Pi_{\rm OPE+DV}(s)$ on the circle of radius  $s_0=m_\tau^2$, for various truncation orders of the OPE expansion. For  $N_{\rm OPE}>13$ the expansion starts to diverge. \label{tab:d0}}\vspace{0.1cm}
\renewcommand{\tabcolsep}{0.4pc} 
\renewcommand{\arraystretch}{1.1} 
\begin{tabular}{ l cc|c c}\hline\hline
$N_{\rm OPE}$ & $\delta^{\rm OPE}_{\rm exact}$ & $\delta^{\rm OPE}_0$ & $\delta^{\rm OPE+DV}_{\rm exact}$&  $\delta^{\rm OPE+DV}_0$ \\\hline
3 &1.98$\times 10^{-2}$  & 6.90$\times 10^{-3}$  & 9.20$\times 10^{-6}$ & 8.73$\times 10^{-6}$ \\
5 &1.98$\times 10^{-2}$  & 6.89$\times 10^{-3}$  & 1.33$\times 10^{-6}$ & 1.14$\times 10^{-6}$ \\
10 &  1.98$\times 10^{-2}$  & 6.89$\times 10^{-3}$ & 1.97$\times 10^{-7}$ &1.34$\times 10^{-7}$  \\
13 &  1.98$\times 10^{-2}$  & 6.89$\times 10^{-3}$ & 1.38$\times 10^{-7}$ &9.31$\times 10^{-8}$  \\
20 & 1.98$\times 10^{-2}$ & 6.89$\times 10^{-3}$  & 3.97$\times 10^{-7}$ &2.99$\times 10^{-7}$\\
30 &  1.98$\times 10^{-2}$ & 6.88$\times 10^{- 3}$ & 6.90$\times 10^{-5}$ & 6.85$\times 10^{-5}$\\
35 &  1.90$\times 10^{-2}$ & 6.62$\times 10^{-3}$ & 3.96$\times10^{-3}$  &  3.95$\times 10^{-3}$ \\
40 &  0.6918  & 0.6908 & 0.6918 & 0.6908\\\hline\hline
\end{tabular}
\end{table}

\begin{table}[hbtp]
\caption{As in Table~\ref{tab:d0} for $s_0=1.5\ {\rm GeV}^2$. The divergent character of the expansion becomes manifest at a lower order, for $N_{\rm OPE}> 7$. \label{tab:d01}}\vspace{0.1cm}
\renewcommand{\tabcolsep}{0.4pc} 
\renewcommand{\arraystretch}{1.1} 
\begin{tabular}{l c c|c c}\hline\hline
$N_{\rm OPE}$ & $\delta^{\rm OPE}_{\rm exact}$ & $\delta^{\rm OPE}_0$ & $\delta^{\rm OPE+DV}_{\rm exact}$&  $\delta^{\rm OPE+DV}_0$ \\\hline
3 &4.76$\times 10^{-2}$  & 2.67$\times 10^{-2}$  & 2.20$\times 10^{-4}$ & 1.81$\times 10^{-4}$ \\
5 &4.75$\times 10^{-2}$  & 2.67$\times 10^{-2}$  & 1.23$\times 10^{-4}$ & 1.01$\times 10^{-4}$ \\
7 &4.76$\times 10^{-2}$  & 2.67$\times 10^{-2}$  & 1.15$\times 10^{-4}$ & 7.98$\times 10^{-5}$ \\
10 & 4.76$\times 10^{-2}$  & 2.67$\times10^{-2}$ & 2.45$\times 10^{-4}$ &1.63$\times 10^{-4}$  \\
15 & 4.76$\times 10^{-2}$ & 2.68$\times10^{-2}$  & 2.23$\times 10^{-3}$ &2.09$\times 10^{-3}$\\
20 &  0.2449 & 0.2263 & 0.2272 & 0.2259 \\\hline\hline
\end{tabular}
\end{table}

 In Table~\ref{tab:d01} we give the results obtained for $s_0=1.5\ {\rm GeV}^2$. The pattern is similar, but the values of both $\delta_{\rm exact}$ and the  lower bound $\delta_0$ are larger. In this case, the divergent character of the expansion starts manifesting itself beyond $N_{\rm OPE} =7$.
 
\section{Strength of the duality violating term}\label{sec:dv}

In this section we shall argue that the quantity $\delta_0$ may be a useful tool for testing  models of DVs in perturbative QCD. We first note that $\delta_0$ quantifies in a certain sense the ``non-analyticity" of the input function $h$ defined in terms  of the spectral function measured at low energies and some chosen theoretical approximant along the circle in the complex plane, such as the OPE or the OPE plus DVs. If  $\delta_0$ is large, the function $h(z)$ defined by this input is far from the class of analytic functions $g(z)$. On the other hand, low values of $\delta_0$ indicate the existence of functions $g$ which are close to $h$.  We can further speculate that, if $\delta_0$ is small, also the difference $\delta_{\rm exact}$ between the physical function and the approximant will be small. Tables~\ref{tab:d0} and \ref{tab:d01} confirm that $\delta_0$ follows quite closely the values of the exact difference. Therefore, we expect the particular approximant that leads to small values of $\delta_0$ as being favored by the ``experimental" input. 

\begin{figure}[htbp]
\centering
\vspace{0.5cm}
\includegraphics[width=2.7in]{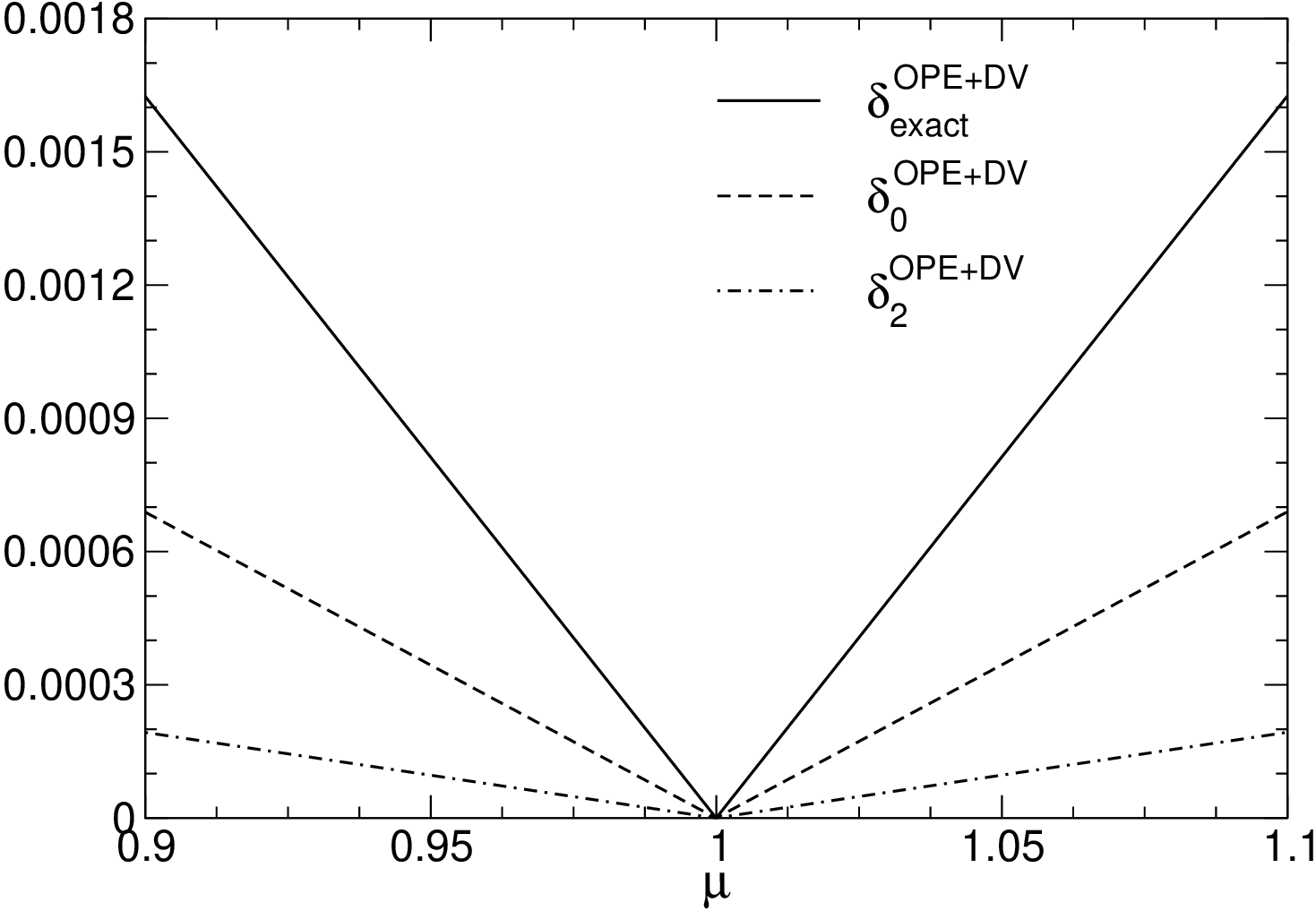}
\caption{Dependence on  $\mu$ of the quantities $\delta_{\rm exact}$ and  $\delta_0$ calculated for $s_0=m_\tau^2$. For comparison 
we present also the quantity $\delta_2$ defined in Eq.~(\ref{delta2}). \label{fig:3}}\vspace{0.3cm}
\end{figure}
In order to  check  this expectation we investigated the sensitivity of the quantities  $\delta_0$ and $\delta_{\rm exact}$ to the magnitude of the DV term added to the OPE expansion in the model investigated in the previous section. As a simple exercise, we introduced a strength parameter $\mu$ multiplying the DV term, \ie, we replaced Eq.~(\ref{OPEDV}) by:
\begin{equation}\label{OPEDV1} \Pi_{\rm OPE+DV}(s)=\Pi_{\rm OPE}(s)+ \mu \,\Pi_{\rm DV}(s)\,.
\end{equation}
From Eq.~(\ref{four1}) it follows that the coefficients~(\ref{cntot}) are replaced by
\beq\label{cntotm}
c_n(\mu)=c_{n}^{\rm \sigma}+ c_{n}^{\rm OPE}+\mu \, c_{n}^{\rm DV}\,,\quad\quad n\ge 1\,,
\eeq
and $\delta_0$ calculated as the norm~(\ref{hank}) of the Hankel matrix will now be a function of the parameter $\mu$.
 
In Fig.~\ref{fig:3} we show the variations  of  $\delta_{\rm exact}^{\rm OPE+DV}$ and  
$\delta_0^{\rm OPE+DV}$ with the parameter $\mu$, taking  as before $N_{\rm OPE}=5$ and $s_0=m_\tau^2$.  
We note that the  quantity $\delta_0$, which can be computed, in principle, from experimental information available for the QCD correlators, has a behavior similar to that of the exact difference $\delta_{\rm exact}$: both exhibit a sharp minimum  at the true value $\mu=1$. For comparison, we show also the quantity $\delta_2$ defined in Eq.~(\ref{delta2}).   In 
 Fig.~\ref{fig:4}  we show the three curves for the circle of radius $s_0=1.5\ {\rm GeV}^2$. 
 We observe that both $\delta_0$ and $\delta_{\rm exact}$  exhibit a small plateau near the minimum. As for the quantity $\delta_2$, it stays below $\delta_0$ in agreement with the exact inequality~(\ref{l2}), and is much less sensitive to the variation of the strength parameter $\mu$.

\begin{figure}[htbp]\vspace{0.5cm}
\includegraphics[width=2.7in]{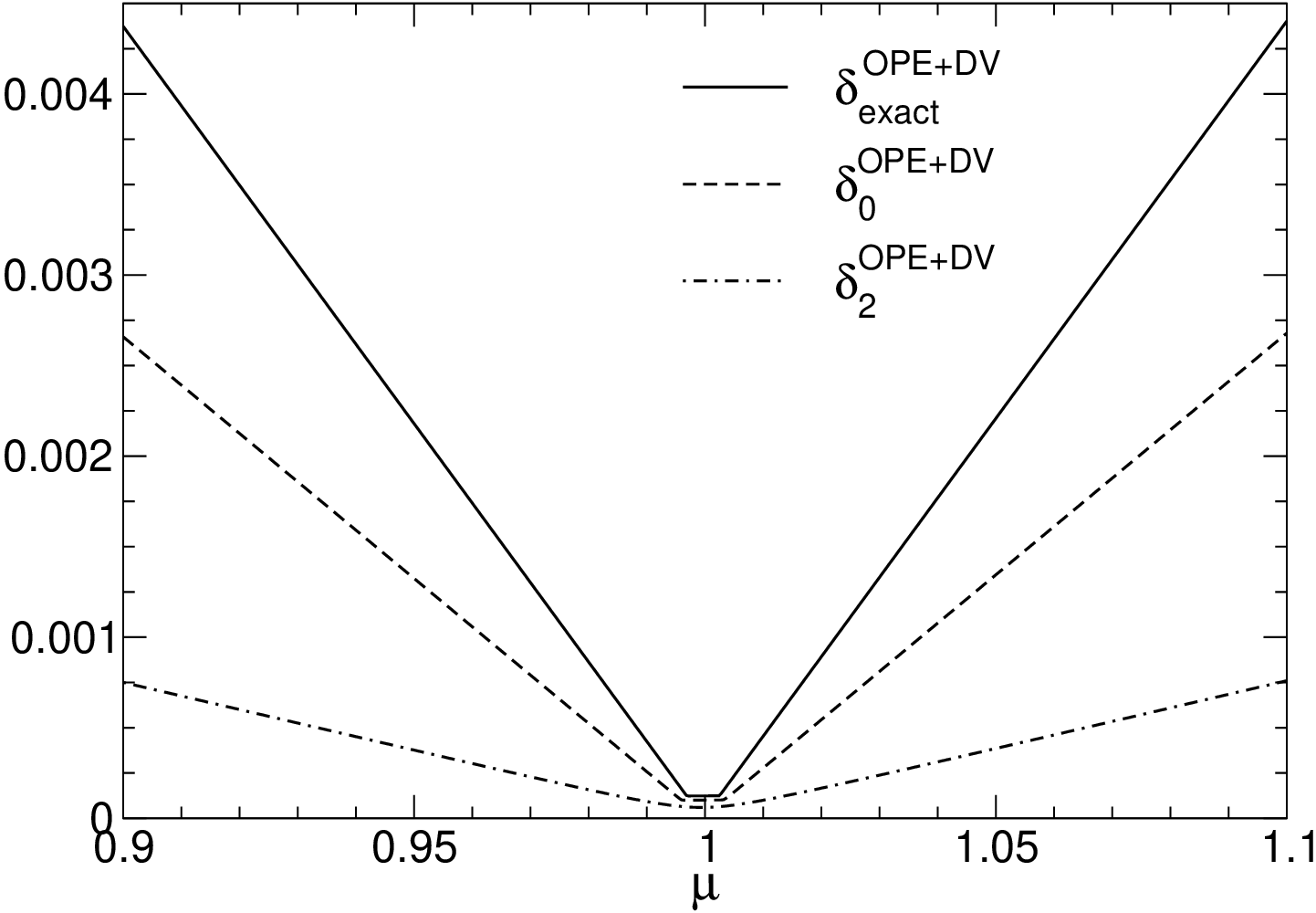}
\caption{As in Fig.~\ref{fig:3} for $s_0=1.5\ {\rm GeV}^2$.\label{fig:4}}
\end{figure}

\section{Discussion}\label{sec:disc}
Within the context of finite-energy sum rules,
one exploits analyticity in order to relate QCD predictions   to physical
measurements  by considering a Cauchy integral along the closed contour shown in Fig.~\ref{fig:0}. Since the exact polarization function is analytic inside the contour, it satisfies the relation 
\beq\label{Cauchy0}
\frac{1}{ 2\pi i}\oint \phi(s)\Pi(s)d s=0\,,
\eeq
where $\phi(s)$ is an arbitrary function holomorphic in the region
$\vert s\vert\le s_0$. This relation can be written alternatively as
\begin{equation}\label{Cauchy}
\frac{1}{\pi}\int\limits_{0}^{s_0}\phi(s) \sigma(s) ds+\frac{1}{ 2\pi i}
\oint\limits_{\vert s\vert=s_0}\phi(s)\Pi(s)d s=0\,,
\end{equation}
in terms of the spectral function $\sigma(s)$ defined in Eq.~(\ref{sigma}). 
 In particular,  choosing 
\beq\label{phi}
\phi(s)=\frac{s^{n-1}}{s_0^{n}}\,, \quad\quad n=1,\ 2,\ \ldots
\eeq
one obtains from Eq.~(\ref{Cauchy0})  the relations
\beq\label{cn0}
c_n=0, \quad\quad n= 1,\ 2,\ \ldots\,,
\eeq
 where $c_n$ are precisely the Fourier coefficients defined in Eq.~(\ref{four1}), but where
 we replace the approximant $\Pi_{\rm QCD}$ by the exact $\Pi_{\rm model}$ and use
 the exact $\sigma_{\rm model}$  calculated from Eq.~(\ref{model0}).   This, of course, 
 yields $\delta_0=0$.

However, in the example presented in Sec.~\ref{sec:model},  we used instead of 
the exact $\Pi_{\rm model}(s)$ on the circle $|s|=s_0$ its  approximants $\Pi_{\rm OPE}$ or $\Pi_{\rm OPE+DV}$. Then the coefficients $c_n$ are all different from 0. Through the relations (\ref{hank}) and (\ref{delta0}), they produce a nonzero $\delta_0$, which measures the non-analyticity of the input.

In  QCD, the asymptotic expansions of the correlators contain a purely perturbative (dimension-0) part, which has been calculated to order $\alpha_s^4$ \cite{BCK08}, and power corrections containing  nonzero vacuum condensates multiplied by logarithmically varying coefficients calculated perturbatively \cite{SVZ}. At the present stage of theoretical knowledge, the power corrections consist of  a limited number $N_{\rm OPE}$   of integer powers of $1/s$ with almost constant coefficients. So, unlike in the model considered above, where most of the individual terms in the OPE contribute to a large numbers of moments, in QCD  each power correction contributes  only to  a definite coefficient $c_n$.  Therefore, for the determination of the condensates it is reasonable to use the conditions
\beq\label{FESR}
c_n =0, \quad\quad n=1,2,...,\, N_{\rm OPE}\ ,
\eeq
defining the so-called ``moment finite-energy sum rules." To further optimize the extraction of the parameters of interest (the strong coupling $\alpha_s$ and the condensates), it is useful to work with the so-called ``pinched" moments, defined by using  weight functions $\phi(s)$ that vanish near the point $s_0$, thus suppressing the contribution of the DV terms \cite{DiPi,DVsup}.

However, when the problem is to discriminate between possible forms for the duality-violating contributions, it is convenient to use weight functions that do not vanish at $s=s_0$. 
In recent phenomenological analyses \cite{Boito1, Boito2},  the extraction of the  parameters 
$\{\vec{p}\}$ entering the OPE and a possible \ansatz\ for DVs was based on a ``fit quality"  $\chi^2$ defined as\footnote{Our discussion here is a simplification of the actual analysis
performed in Refs.~\cite{Boito1, Boito2}.}
\beq\label{chi2}
\chi^2(\vec{p}) =\sum_{n\ge 1} \frac{c_n^2(\vec{p})}{\epsilon^2_n}\,,
 \eeq 
where
\beq\label{epsn}
 \epsilon_n=\delta c_n^\sigma
\eeq 
are the errors of the experimental moments. If the errors on the $c_n^\sigma$ are 
uncorrelated, this fit quality is the usual $\chi^2$. 
The best values of  $\{\vec{p}\}$ are found from the maximization  of  a likelihood function in the parameter space, defined as
${\cal L}_2(\vec{p})\sim \exp[-\chi^2(\vec{p})/2]$. This approach generalizes the strict sum rules~(\ref{FESR}), 
allowing for fluctuations within the experimental errors $\epsilon_n$ of the coefficients $c_n$ responsible for non-analyticity.

 One may view the above likelihood as being induced by the analyticity properties known to be satisfied by the physical function. More formally, we note that $\chi^2$ can be compared to the quantity $\delta_2$ of Eq.~(\ref{delta2}).  If we define a modified version $\bar \delta_2(\vec{p}) $ of $\delta_2(\vec{p})$ by replacing $c_n$ in Eq.~(\ref{delta2}) by
 \beq\label{cnbar}
\bar c_n(\vec{p})= \frac{c_n(\vec{p})}{\epsilon_n}\,,
\eeq
 \ie, the original  coefficients ${c_n(\vec{p})}$ normalized by the experimental errors, 
 we see that
\beq\label{chi2L2}
\chi^2(\vec{p}) \equiv \bar \delta_2^2(\vec{p})\,.
\eeq
Therefore, in this sense, $\chi^2(\vec{p})$ quantifies the ``non-analyticity," defined 
here as the minimal distance measured by
 the $L^2$ norm on the set $H^2$, of a function  $h(\zeta)$ having as negative frequency coefficients in the expansion~(\ref{hpm}), instead of the coefficients $c_n$,  the normalized ratios~(\ref{cnbar}).

\begin{figure}[t]
\centering
\vspace{0.5cm}
\includegraphics[width=2.7in]{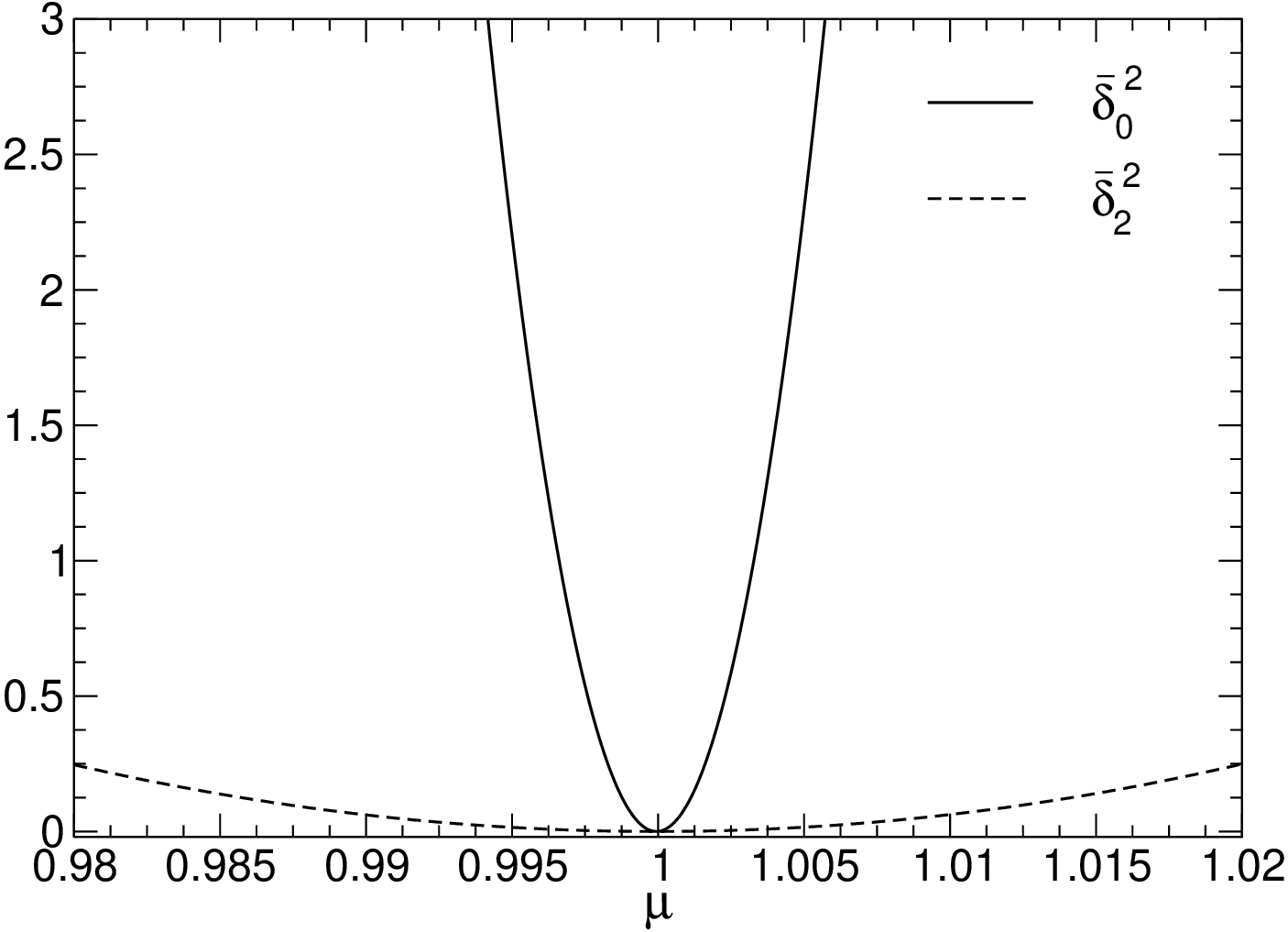}
\caption{Variation with $\mu$ around the exact value of the quantities $\bar\delta_0^2$ and  $\bar\delta_2^2$, calculated for $s_0=m_\tau^2$.  \label{fig:6}}\vspace{0.3cm}
\end{figure}

Using the comparison between the minimization problems based on the $L^\infty$ norm or the $L^2$ norm discussed in Sec.~\ref{sec:solution}, one might also consider the quantity 
\beq\label{delta0bar}
\bar \delta_0^2(\vec{p}) = \Vert \bar{\cal H}(\vec{p})\Vert^2\,,
\eeq
\ie, the norm squared of a Hankel matrix   $\bar{\cal H}$  constructed  from the coefficients $\bar c_n(\vec{p})$ by a relation similar to Eq.~(\ref{hank}):
\begin{equation}\label{hankbar}
\bar{\cal H}_{nm}(\vec{p})=\bar c_{n+m-1}(\vec{p})\,, \quad \quad n,m\geq 1\,.
\end{equation}
We see that $\bar \delta_0^2(\vec{p})$ quantifies  the non-analyticity of the same function $h$ considered in the $\chi^2$ test, but  measured now in the $L^\infty$ norm. As above,  we may postulate a likelihood ${\cal L}_0(\vec{p})\ \sim \exp[-\bar \delta_0^2(\vec{p})/2]$ in the space of parameters,  
and determine the best parameters from the minimum of $\bar \delta_0^2(\vec{p})$. Since the $L^\infty$ norm tests local properties of the modulus of the function, while the $L^2$ norm tests the magnitude of the modulus only in the average,  we might expect the new test to be stronger and have a greater sensitivity to the parameters that control the non-analyticity of the  theoretical input.

 In order to illustrate this idea we considered again the model discussed in Sec.~\ref{sec:model} and generated  errors $\epsilon_n$ on the coefficients $c_n^\sigma$ by varying the input spectral function.
 The experimental data from $\tau$ hadronic decays \cite{Aleph,Aleph1, Opal} have in general larger (relative) errors in the low energy region and near the upper  limit $s=m_\tau^2$, and smaller errors in the intermediate region. This implies larger errors $\delta c_n^\sigma$ for the higher order coefficients $c_n^\sigma$, which are dominated by the large energy region. To simulate this  situation,  we assumed errors on $\sigma_{\rm model}(s)$ of 10\%  for $\sqrt{s}$ below 0.5 ${\rm GeV}$ and above 1.7 ${\rm GeV}$, and of 3\% from 0.5 to 1.7 ${\rm GeV}$.  The errors on $c_n^\sigma$ were obtained assuming fully correlated errors for $\sigma(s)$ at different energies. The errors $\epsilon_n$ thus obtained  increase with $n$, reaching the level of 10\% for $n$ around 50.    Since this exercise is only for illustrative
 purposes, we ignore the correlations between $c_n^\sigma$ and $c_m^\sigma$ for $n\ne m$.

We then used the quantities $\bar \delta_2^2$ and $\bar \delta_0^2$ defined in Eqs.~(\ref{chi2L2}) and (\ref{delta0bar}), respectively, and checked their sensitivity to the strength parameter $\mu$ introduced in Eq.~(\ref{OPEDV1}). In both cases we have used the same number of coefficients, $N=500$.
 As shown in Fig.~\ref{fig:6}, the variation with $\mu$ of  $\bar \delta_0^2(\mu)$ is much more rapid than that of  $\bar \delta_2^2(\mu)$.  
 This figure shows how also $\bar \delta_0^2(\mu)$ can 
 in principle be used as a potentially interesting fit quality.

As mentioned above, in this exercise we neglected correlations among the coefficients $c_n^\sigma$, while in practice one expects these coefficients  to be
 correlated, since they are all computed from the same spectral function. 
The definitions of $\bar \delta_2^2(\vec{p})$  in
 Eqs.~(\ref{chi2}--\ref{chi2L2}) and of $\bar\delta_0^2(\vec{p})$ in  Eq.~(\ref{delta0bar}) ignore correlations.
Moreover, unlike for the case of $\bar \delta_2^2(\vec{p})$,
it is not immediately clear how to incorporate them in the $L^\infty$-norm based test.   One could, of course, estimate the error
on the parameter $\mu$ (or, in general, the fit parameters $\vec p$) by 
Monte Carlo, taking the full covariance matrix into account. It is of interest to find out which of the two quantities, $\bar \delta_2^2(\vec{p})$
or $\bar \delta_0^2(\vec{p})$, used in this way, would lead to a smaller error.
This is beyond the scope of the present article.

We end this section with the observation that both 
 quantities  $\bar \delta_2^2$ and $\bar \delta_0^2$ depend quadratically on the coefficients $c_n$. While this is obvious for the $L^2$ norm,  the dependence of  $\bar \delta_0^2$ on the $c_n$ is not so transparent. We note however that from the definition~(\ref{Hinf}) of the  $L^\infty$ norm we expect $\delta_0$ to exhibit  an almost piecewise linear dependence  on the coefficients $c_n$ and consequently on the parameter  $\mu$. This feature is confirmed in Figs.~\ref{fig:3} and \ref{fig:4}. The quantity $\bar \delta_0^2$, defined as the square of $\bar\delta_0$, is therefore expected to have a piecewise quadratic   dependence on the coefficients $c_n$. 
 Numerical tests show that the dependence of $\bar \delta_0^2$ on the higher order coefficients $c_n$ is indeed very close to quadratic. The resulting near-quadratic dependence of $\bar \delta_0^2$ on  $\mu$ is seen in Fig.~\ref{fig:6}.

\section{Conclusions and outlook}\label{sec:out}

In this article we have argued that the problem of violations of quark-hadron duality
in QCD can be investigated with the methods of functional analysis.
We showed that the distance, measured in the $L^\infty$ norm along a contour in the complex plane, between an exact QCD correlation function and its theoretical approximation by the OPE plus possible DV terms, must be larger  than a certain  calculable quantity $\delta_0$. This quantity is defined by a  functional minimization problem, which was solved by a duality theorem in functional optimization. As shown in Sec.~\ref{sec:solution}, the problem can be reduced to a numerical algorithm. This allows the calculation of $\delta_0$ as the norm of a Hankel matrix constructed in terms of the Fourier coefficients~(\ref{four}), which can be decomposed as in Eq.~(\ref{four1}).

We demonstrated the usefulness of the quantity  $\delta_0$ by studying a model proposed in Ref.~\cite{Cata2}, which allows for a study of the interplay between an asymptotic OPE
and duality violations. In this model,
we showed that $\delta_0$ is smaller than but quite close to the exact distance, which can also be calculated in this case, but is unknown in a
realistic physical situation. This makes  $\delta_0$  a potentially suitable tool for the investigation of duality violations, for which an analytic form is not known in general.  In particular, we showed that $\delta_0$ is very sensitive to the variation of the parameter $\mu$ introduced in Eq.~(\ref{OPEDV1}) to measure the strength of the duality violating contribution.

We note that with the introduction of the parameter $\mu$, the problem becomes analogous to 
the search for new physics beyond the Standard Model in experiments at very high energies.  There, one tests for the presence of new physics through the ``strength parameter'' $\mu$ of the signal, while treating the Standard Model as background (for likelihood-based statistical tests used for the discovery of new phenomena see, for instance, Ref.~\cite{Cowan}).
In our case, we want to detect the presence of DVs (the ``new physics'') in addition to the OPE
(the ``known physics''),  which depend on a set of parameters $\vec{\theta}$, often
referred to as  ``nuisance parameters."

In Sec.~\ref{sec:disc}, we introduced modified versions $\bar\delta_0$ and
$\bar\delta_2$ of $\delta_0$ and a similar
distance $\delta_2$ measured in the $L^2$ norm, to account for the fact that in a realistic
application the Fourier coefficients~(\ref{four}) will only be known within certain errors.
We showed how $\bar\delta_0$ can be computed from a modified Hankel matrix, and
we compared the modified quantities to the usual $\chi^2$ commonly used to quantify
the distance between data and theory.   The standard $\chi^2$ can be interpreted as
$\bar\delta_2^2$, the minimal distance squared, but measured in the $L^2$ norm
instead of the $L^\infty$ norm.
The discussion in Sec.~\ref{sec:disc} suggests that also $\bar \delta_0^2$ can 
 in principle be used as a potentially interesting fit quality, but a practical implementation remains to be explored.

\subsection*{Acknowledgements} IC acknowledges support from the Ministry of Education under Contract PN No 09370102/2009  and from UEFISCDI under contract Idei-PCE No 121/2011,
MG is supported in part by the US Department of Energy
and SP is supported by CICYTFEDER-FPA2011-25948, SGR2009-894, and
the Spanish Consolider-Ingenio 2010 Program CPAN (CSD2007-00042).
This work was initiated during the Workshop on {\em Hadronic contributions to muon g-2}, Mainz, 1-5 April 2014.

\end{document}